\documentclass[twocolumn,prd,superscriptaddress,preprintnumbers,nofootinbib]{revtex4}[11pt]
\pdfoutput=1
\usepackage{amsmath,amssymb,graphicx}
\graphicspath{{figs/}}
\usepackage{epsf,verbatim}
\usepackage{hyperref}
\usepackage{comment}
\usepackage{color}
\usepackage{slashed}
\usepackage{subfigure}
\usepackage[usenames,dvipsnames]{xcolor}
\usepackage{comment}
\usepackage{mdwlist, paralist}
\usepackage{rotating}
\usepackage{multirow}

\usepackage[utf8]{inputenc}

\def\to{\rightarrow}

\def\mL{\mathcal{L}}

\def\tbf{\textbf}

\emergencystretch=\maxdimen
\begin{document}

\title{Composite Higgs Meets Planck Scale: Partial Compositeness from Partial Unification}

\preprint{NCTS-PH/1907}

\author{Giacomo Cacciapaglia}
\thanks{g.cacciapaglia@ipnl.in2p3.fr}
\author{Shahram Vatani}
\thanks{vatani@ipnl.in2p3.fr}
\affiliation{Institut de Physique des 2 Infinis (IP2I), CNRS/IN2P3, UMR5822, F-69622 Villeurbanne, France}
\affiliation{Universit\'e de Lyon, France; Universit\'e Claude Bernard Lyon 1, Lyon, France}

\author{Chen Zhang}
\thanks{czhang@cts.nthu.edu.tw}
\affiliation{Physics Division, National Center for Theoretical Sciences, Hsinchu, Taiwan 300}

\begin{abstract}
Providing an Ultra-Violet completion valid up to the Planck scale is of paramount importance to validate the composite Higgs paradigm, at par with supersymmetry.
We propose the first complete and feasible framework, based on partial unification of a confining hypercolor gauge group, where couplings of the standard model fermions are mediated by both gauge and scalar bosons. We demonstrate our approach by providing an explicit model based on a Techni-Pati-Salam unification, $SU(8)_{\rm PS}\times SU(2)_L\times SU(2)_R$, able to generate masses for all fermion generations, including neutrinos, via partial compositeness. We predict an $Sp(4)$ hypercolor group, and lattice studies will be crucial to validate the model.
\end{abstract}

\maketitle

\setcounter{equation}{0} \setcounter{footnote}{0}


The mechanisms behind electroweak symmetry breaking (EWSB) and fermion mass generation
are of crucial importance to our quest for the fundamental laws of nature, yet
in the Standard Model (SM) they are simply parameterized via the vacuum expectation
value of an elementary scalar~\cite{Higgs:1964pj}.
The extreme economy of the SM Higgs sector,  however, precludes
a deeper understanding of how the electroweak scale is stabilized against radiative corrections
and how fermion masses and flavor mixing are generated.
A time-honored alternative is the idea that the Higgs may emerge as
a composite pseudo-Nambu-Goldstone boson (pNGB) associated with the spontaneous global symmetry breaking
in a new strong dynamics sector~\cite{Kaplan:1983fs,Contino:2010rs}.
Such a pNGB Higgs could naturally behave like the SM Higgs if a hierarchy between the electroweak scale
$v = 246$~GeV and the compositeness scale $f \approx \mathcal{O} ({\rm few})$~TeV is generated~\cite{Agashe:2005dk}.
So far, this approach has been studied in an effective field theory approach, with no successful attempt
to construct a theory whose validity can be extended up to the Planck scale. Only within such extensions, composite Higgs
models can be considered as serious alternatives to the SM, at par with the most widely studied supersymmetric scenarios.
In this letter, we propose the first complete feasible extension of this type.

The most challenging
aspect in this framework is the generation of large enough masses for the SM fermions, especially
the top quark, without contradicting flavor constraints. This
typically requires a conformal, or walking, phase right above the confinement scale in order to separate the
scale of flavor symmetry breaking~\cite{Holdom:1981rm} from the condensation scale. The
partial compositeness (PC) paradigm~\cite{Kaplan:1991dc}, which postulates a linear mixing between the SM fermions and operators
in the strong sector, provides a promising solution at an effective theory level. Nevertheless,
Ultra-Violet (UV) completions of PC turn out to be very challenging. There exist
elegant constructions based on extra dimensions~\cite{Contino:2003ve,Agashe:2004rs,Hosotani:2005nz}, which are however not UV-complete~\cite{Gies:2003ic,Morris:2004mg,Panico:2015jxa}.
Instead, we stick to four-dimensional (4D) quantum field theories describable by a Lagrangian~\cite{Barnard:2013zea,Ferretti:2013kya,Vecchi:2015fma},
which have the virtue of being amenable to lattice studies~\cite{Bennett:2017kga,Ayyar:2018zuk,Ayyar:2018glg,Bennett:2019jzz,Hasenfratz:2016gut}.

In our 4D setup, the sector of strong dynamics features a new confining gauge group, hypercolor (HC), and a
set of new fermions (hyperfermions) charged under it.
PC is thus realized via four-fermion operators involving one SM fermion
and three hyperfermions. Remarkably, all possible combinations of HC groups and hyperfermion representations can be enumerated,
by considering the minimal setting
in which hyperfermions are, at most, in two distinct irreducible representations of a
simple HC group~\cite{Ferretti:2013kya}.
The number of solutions is essentially finite once we
take into account various constraints, especially the requirement that the HC theory at low energy
is outside the conformal window~\cite{Ferretti:2016upr,Belyaev:2016ftv}, i.e. it confines. These theories are, however, {\it not} UV complete,
as they only accommodate for top PC and do not explain the origin of the four-fermion  interactions.
In this letter, we aim at offering a first truly UV complete realization of this class of models, which includes
renormalizable interactions all the way up to the Planck scale and for all SM generations.
We note that alternative ideas have been put forward, based on UV safety~\cite{Cacciapaglia:2018avr},
on global symmetries~\cite{Cacciapaglia:2019vce}, on the presence of light HC-ed scalars~\cite{Sannino:2016sfx}
or on supersymmetric dualities~\cite{Caracciolo:2012je,Marzocca:2013fza}. Yet, no complete
nor fully realistic proposal exists in the literature.

There are two crucial questions that need to be answered in this setup.
First, large anomalous dimensions of three-fermion operators are needed to generate a large
enough top Yukawa coupling~\cite{Contino:2010rs}. This, in turn, requires that the HC theory enters
a strongly-coupled near-conformal regime between the confinement scale and the scale $\Lambda_F$
where flavor physics is generated. The latter needs to be large enough to evade flavor constraints.
Whether this is possible or not, it can only be established by intrinsically non-perturbative methods, e.g. lattice simulations,
while perturbative calculations can only give a qualitative indication~\cite{Pica:2016rmv,BuarqueFranzosi:2019eee}.

Second, the four-fermion couplings are a
valid description only up to the scale $\Lambda_F$ where they need to be replaced by a mediator.
It is most straightforward to envision two extreme possibilities for UV completions. The first is to consider a chiral gauge theory with
only fermionic matter content, in which the full gauge group experiences a series of self-breaking
to a low-energy model containing the SM and HC gauge groups with appropriate fermion field
content. The massive gauge bosons appearing in the self-breaking process could act as mediators
for the PC four-fermion operators. However, this approach is very similar to
 Extended Technicolor~\cite{Eichten:1979ah}, of which type, despite intense efforts~\cite{Hill:2002ap,Appelquist:2003uu,Appelquist:2003hn,Appelquist:2004ai}, no fully realistic
models have been constructed so far. The second, is to introduce new elementary scalar mediators, as in bosonic Technicolor~\cite{Samuel:1990dq}.
The latter approach, however, would not buy any advantage with respect to the SM Higgs sector, in terms
of radiative stability or understanding of the flavor origins.

In this letter, we propose a class of models that stand in between these two extremes. While we keep the ambitious
goal of building a fully realistic model, we are modest in that high-scale elementary scalars are allowed
to exist. The crucial ingredient in our construction is the {\it partial unification} (PU) of HC with other SM gauge
groups.~\footnote{In contrast to Extended Technicolor, we do not require unification with family symmetries.} The
massive vectors from gauge symmetry breaking, together with high-scale scalars, act as PC mediators.
Compared to pure scalar mediation, the PU ensures less arbitrariness as a somewhat limited number
of interactions are able to generate {\it automatically} all the necessary four-fermion couplings. In the following,
we will present an explicit model and discuss its features, including testable correlations in the low-energy model.

The most non-trivial issue in our PU approach is finding the appropriate field content that, once decomposed
under the SM and HC groups, contains the SM fermions and the needed hyperfermions,  without any gauge
anomalies. There is, a priori, no guarantee that such a solution exists for all low-energy models in refs~\cite{Ferretti:2016upr,Belyaev:2016ftv}.
To provide a first example, we base our construction on simplicity: we try to embed the SM quarks (top) and
the hyperfermions generating the Higgs in the same multiplets. To do so, it suffices to unify the QCD color and HC
as block-diagonal subgroups of a unified group $\mathcal{G}_{\rm EHC}$ (extended-HC). Scrolling the gauge groups
in the models of ref.~\cite{Ferretti:2013kya}, we see that this construction is only possible for models based on $Sp(4)_{\rm HC}$,
like model M8 in the nomenclature of ref.~\cite{Belyaev:2016ftv}.
The template of this model contains 4 hyperfermions in the fundamental $\textbf{F}$, transforming as a doublet of $SU(2)_L$
and a doublet of $SU(2)_R$, and at least 6 in the two-index anti-symmetric $\textbf{A}_2$. The latter ones carry QCD charges
and are needed to generate colored PC operators. Their quantum numbers will be predicted in our model. The first PU unification
step would thus consist in unifying $Sp(4)_{\rm HC} \times SU(3)_c \subset SU(7)_{\rm EHC}$, embedding the four $\textbf{F}$
with the top and bottom quark fields. To include leptons in the model, we follow the same quark-lepton unification
mechanism of the Pati-Salam model~\cite{Pati:1974yy}, and extend $SU(7)_{\rm EHC} \to SU(8)_{\rm PS} \times SU(2)_R$.
The final model, therefore, has gauge symmetry
\begin{equation}
\mathcal{G}_{\rm TPS} = SU(8)_{\rm PS} \times SU(2)_R \times SU(2)_L\,,
\end{equation}
and we dub it \emph{Techni-Pati-Salam model} (TPS). The field content of TPS is summarized in Table~\ref{table:sffc},
where one can recognize the SM quarks and leptons, and the Higgs hyperfermions, embedded in bi-fundamentals of $SU(8)_{\rm PS}$ and of the two $SU(2)$'s. The fermion singlet $N$ is introduced to explain the small neutrino masses via an inverse see-saw mechanism~\cite{Mohapatra:1986bd}, while $\Xi$ is needed to generate PC. The scalars are introduced in order to generate the breaking of the PU gauge groups in the following three steps:
\begin{itemize}
\item[{\it a})] $SU(8)_{\rm PS} \times SU(2)_R \to SU(7)_{\rm EHC} \times U(1)_E$\,,
\item[{\it b})] $SU(7)_{\rm EHC} \to SU(4)_{\rm CHC} \times SU(3)_c \times U(1)_X$\,,
\item[{\it c})] $SU(4)_{\rm CHC} \to Sp(4)_{\rm HC}$, $U(1)_X \times U(1)_E \to U(1)_Y$\,;
\end{itemize}
and to introduce flavor mixing.
Note that the first two steps could take place at the same high scale, close to the Planck mass, while the step {\it c}
can take place at lower energies and be radiatively induced. In the following we will focus on the generation of the
flavour couplings and leave a detailed study of the scalar sector to a further study. More details on this model will be
presented in a forthcoming technical publication~\cite{Giacomo:2019ehd}.

\begin{table}[t!]
\begin{tabular}{|c|c|c|c|c|c|c|}
\hline
Field & Spin & $SU(8)_{\rm PS}$ & $SU(2)_L$ & $SU(2)_R$ & \# & $Q_G$\\
\hline
$\Omega$ & $1/2$ & $\textbf{8}$ & $\textbf{2}$ & $\textbf{1}$ & $3$  & $1$\\
\hline
$\Upsilon$ & $1/2$ & $\bar{\textbf{8}}$ & $\textbf{1}$ & $\textbf{2}$ & $3$ & $-1$\\
\hline
$\Xi$ & $1/2$ & $\textbf{70}(=\textbf{A}_4)$ & $\textbf{1}$ & $\textbf{1}$ & $1$ & $0$\\
\hline
$N$ & $1/2$ & $\textbf{1}$ & $\textbf{1}$ & $\textbf{1}$ & $3$ & $0$\\
\hline
$\Phi$ & $0$ & $\textbf{8}$ & $\textbf{1}$ & $\textbf{2}$ & $1$ & $1$\\
\hline
$\Theta$ & $0$ & $\textbf{28}(=\textbf{A}_2)$ & $\textbf{1}$ & $\textbf{1}$ & $2$ & $2$\\
\hline
$\Delta_R$ & $0$ & $\textbf{56}(=\textbf{A}_3)$ & $\textbf{1}$ & $\textbf{2}$ & $1$ & $1$\\
\hline
$\Delta_L$ & $0$ & $\textbf{56}(=\textbf{A}_3)$ & $\textbf{2}$ & $\textbf{1}$ & $1$ & $1$\\
\hline
$\Psi$ & $0$ & $\textbf{63}(=\textbf{Adj})$ & $\textbf{1}$ & $\textbf{1}$ & $2$ & $0$\\
\hline
\end{tabular}
\caption{Scalar and (left-handed Weyl) fermion field content. $\textbf{A}_n$ and $\textbf{Adj}$ denote
the $n$-index anti-symmetric and the adjoint representations, respectively.
The second-to-last column indicates the multiplicities, while the last the global $U(1)_G$ charges.
\label{table:sffc}}
\end{table}

For convenience, we adopt the following notations for displaying representations at the HC level, e.g.
\begin{align}
(\textbf{4},\textbf{3})_{1/6} & \Rightarrow (Sp(4)_{\rm HC},SU(3)_c)_{U(1)_Y}\,.
\end{align}
The SM fermions are embedded in the $\Omega$'s and $\Upsilon$'s as
\begin{align}
\Omega^p & =\left((L_u^p,\ u_L^p,\ \nu_L^p)_{\frac{1}{2}},\ (L_d^p,\ d_L^p,\ e_L^p)_{-\frac{1}{2}}\right)\,, \\
\Upsilon^p & =\left((U_d^p,\ d_R^{pc},\ e_R^{pc})_{\frac{1}{2}},\ (D_u^p,\ u_R^{pc},\ \nu_R^{pc})_{-\frac{1}{2}}\right)\,,
\end{align}
where $p=1,2,3$ is a family index, and the superscript $c$ indicates the charge-conjugated field.
The subscript $\pm 1/2$ indicates the eigenvalue under the diagonal generator of $SU(2)_L$ for $\Omega$,
and $SU(2)_R$ for $\Upsilon$, of the components.
The hyperfermions  have quantum numbers
\begin{align}
L_{u/d}^p\Rightarrow (\textbf{4},\textbf{1})_0,\; U_d^p\Rightarrow (\textbf{4},\textbf{1})_{1/2},\; D_u^p\Rightarrow (\textbf{4},\textbf{1})_{-1/2}\,.
\end{align}
The {\bf 70} decomposes as
$
\Xi=\begin{pmatrix}
U_u & \chi & \rho & \eta & \omega \\
D_d & \tilde{\chi} & \tilde{\rho} & \tilde{\eta} & \tilde{\omega} \\
\end{pmatrix}
$
with
\begin{align}
U_u & \Rightarrow (\textbf{4},\textbf{1})_{-1/2},\;
\chi\Rightarrow (\textbf{5},\textbf{3})_{-1/3},\;
\rho\Rightarrow (\textbf{1},\textbf{1})_0, \nonumber \\
\eta & \Rightarrow (\textbf{4},\bar{\textbf{3}})_{-1/6},\;
\omega\Rightarrow (\textbf{1},\textbf{3})_{-1/3};
\end{align}
while the quantum numbers of the fields in the bottom row are conjugates of the ones in the upper row
(the two rows form conjugate multiplets of $SU(7)_{\rm EHC}$).
The role of $\Xi$ is to contain the 6 hyperfermions in the $\textbf{A}_2$ representation
of $Sp(4)_{\rm HC}$, $\chi$ and $\tilde\chi$. In addition, it contains two more electroweakly charged
hyperfermions, $U_u, D_d$, and QCD-colored hyperfermions $\eta,\tilde\eta$ that will play a role in
the generation of lepton masses. Finally, $\omega,\tilde\omega$ is a vector-like partner of the right-handed
down quarks, and $\rho$ a neutral singlet.

The scalar fields in Table~\ref{table:sffc} are responsible for the breaking of the PU gauge group:  step {\it a} can be
achieved only by $\Phi$, with vacuum expectation value $v_\Phi$;  {\it b} can be achieved by the adjoint $\Psi$ at a scale $v_\Psi$;
finally {\it c} can be achieved by $\Theta$ at $v_\Theta$.
We anticipate that the second copy of $\Psi$ and $\Theta$, together with the two $\Delta$'s, are required to generate the necessary flavor structure in the model, while $\Delta_R$ could also participate to the breaking steps {\it b} and {\it c}.~\footnote{At the price of breaking baryon number, as defined later in the letter.} We will implicitly assume $v_\Phi \approx v_\Psi \lesssim M_{\rm Pl}$, and $v_\Theta \approx \Lambda_F \lesssim v_\Phi$.

We are now ready to analyze the model in more details: in this letter our goal is to demonstrate that the
model can generate the correct flavor structures and composite dynamics at low energy.
The complete UV Lagrangian at the Planck scale can be decomposed as
\begin{align}
\mL=\mL_{\rm kin}+\mL_{\rm pot}+\mL_{\rm Yuk}\,,
\end{align}
where the first term includes the kinetic terms with gauge interactions, the second a potential for
the scalar fields $\mL_{\rm pot} = - V(\Phi,\Theta,\Delta,\Psi)$, and the third the fermion masses and
Yukawa couplings:
\begin{multline} \label{eq:Yuk}
\mL_{\rm Yuk}  =-\mu_N^{pq}\ N^pN^q  -\mu_\Xi\  \Xi\Xi - \lambda_\Psi^\alpha\  \Xi\Psi^\alpha\Xi \\
   - \left( \lambda_\Phi^{pq} \Upsilon^p\Phi N^q   + \lambda_{\Theta L}^{pq,\alpha}\ \Omega^p\Theta^{\alpha*}\Omega^q + \lambda_{\Theta R}^{pq,\alpha}\ \Upsilon^p\Theta^\alpha\Upsilon^q  \right. \\
   \left. + \lambda_{\Delta R}^p\ \Upsilon^p\Delta_R^\ast\Xi + \lambda_{\Delta L}^p\ \Omega^p\Delta_L\Xi + \text{h.c.} \right)\,,
\end{multline}
where $p,q = 1,2,3$ and $\alpha = 1,2$.
Without loss of generality, we can use the flavor symmetries of the scalars to cast the vacuum expectation values
on $\Theta^1$ and $\Psi^1$, and diagonalise $\mu_N^{pq}$ and the Yukawa couplings $\lambda^{pq,1}_{\Theta L/R}$ in the fermion flavour space. This simple analysis  shows that flavour structures are contained in $\lambda^{pq,2}_{\Theta L/R}$ and in the $\Delta_{L/R}$ Yukawas,
while $\lambda^{pq}_\Phi$ carries flavour structures for neutrinos. Also, the Yukawa sector above allows for a global $U(1)_G$ symmetry, whose charges are listed in Table~\ref{table:sffc}, which is broken both spontaneously and explicitly in the scalar potential. The latter guarantees the absence of massless Goldstones.

To analyze how PC arises, we first study the 3$^{\rm rd}$ generation only, for which all four-fermion couplings
can be generated by gauge-mediators. By examining the index decomposition of gauge-fermion interactions,
we can determine all vector-mediated operators. Below we give some examples after Fierzing, while a complete
list will be presented in ref.~\cite{Giacomo:2019ehd}:
\begin{equation} \label{eq:vmpc4f} \begin{array}{ccc}
(\bar{D}_u^3 \bar{U}_u) (\chi t_R^c)\,, & (\bar{U}_d^3 \bar{U}_u) (\chi b_R^c)\,, & (\bar{L} \bar{D}_d) (\tilde{\chi} q_L)\,, \\
(\bar{D}_u^3 \bar{\tilde{\eta}}) (\chi \nu_{\tau R}^c)\,, & (\bar{U}_u^3 \bar{\tilde{\eta}}) (\chi \tau_R^c)\,, & (\bar{L} \bar{\eta}) (\tilde{\chi} l_L)\,;
\end{array}
\end{equation}
where $L\equiv(L_u^3,L_d^3),q_L\equiv(t_L,b_L),l_L\equiv(\nu_{\tau L},\tau_L)$. The couplings above
allow to identify the PC operators: for instance, the $t_R^c$ partner $T_L = \langle \bar{D}_u^3 \bar{U}_u \chi \rangle$.
We also find that all PC operators for quarks, of which eq.~\eqref{eq:vmpc4f} is a subset, are mediated by
a $(\textbf{4},\textbf{3})_{1/6}$ vector that acquires its mass from the
EHC breaking {\it b}. Similarly, all lepton PC operators involve $\eta$--$\tilde{\eta}$ and are mediated
by a $(\textbf{4},\textbf{1})_{1/2}$ vector that acquires its mass
from PS breaking {\it a}. Thus the difference between the two breaking scales may explain the $t$--$\tau$
mass difference. The tau neutrino is predicted to receive a Dirac mass degenerate with
the charged tau, and for this reason we introduced the singlet $N$ that mixes via the
Yukawa $\lambda_\Phi^{33}$. Note that the Yukawas $\lambda_\Psi^1$ and
$\lambda_{\Theta L/R}^{33,1}$ contribute to the masses of the $\Xi$ components and of the
$L^3_{u/d}$, $U_d^3$, $D_u^3$ hyperfermions. As all of them play an important role in PC, they need
to be much lighter than the PU breaking scale, implying $\lambda_\Psi^1,\ \lambda_{\theta L/R}^{33,1} \ll 1$.
As the scalar-mediated PC couplings can only be proportional to $\lambda_\Psi^\alpha \lambda_{\Theta L/R}^{33,\beta}$
and $(\lambda_{\Delta L/R}^3)^2$, generating sizeable couplings needs either to introduce a second pair $\Theta_{L/R}^2$--$\Psi^2$
with sizeable Yukawas, or the $\Delta_{L/R}$ scalars. A mass split between top and bottom can only be generated by a different
running of the $t_R^c$--$b_R^c$ PC operators, or by the contribution of scalar mediation.

We can now analyse the more general and realistic case with 3 generations.
A first issue we encounter is the presence of too many hyperfermions, as each SM generation carries its own set in the multiplets $\Omega^p$ and $\Upsilon^p$. It may be tempting to let all hyperfermions participate in the EWSB, leading to a multi Higgs doublet model at low energy, while the mass hierarchies are justified by a hierarchy of hyperfermions masses, generated by the (diagonal) $\lambda_{\Theta L/R}^{pp,1}$ couplings.
This scenario, corresponding to the effective  multi-scale model of refs~\cite{Cacciapaglia:2015dsa,Panico:2016ull}, would however be at odds with other required properties of the hypercolor dynamics.
For flavor physics, the dynamics needs to develop
an Infra-Red (IR) near-conformal behavior (walking) in order to generate large enough anomalous dimensions for the PC operators.
We know that $\chi,\tilde{\chi},\eta,\tilde{\eta}$ cannot be much heavier than the condensation scale as they are necessary for quark and lepton PC. The $Sp(4)_{\rm HC}$ theory, therefore, has $6\times\tbf{A}_2$ (accounting for $\chi,\tilde{\chi}$)
and $N_F\times\tbf{F}$ Weyl flavors, where $N_F = 20$ counting all the hyperfermions.
This theory looses asymptotic freedom for $N_F > 20$, thus if all $\tbf{F}$--flavors are light
compared to the scale of $SU(4)_{\rm CHC} \to Sp(4)_{\rm HC}$ breaking, the
theory can at best approach a weakly-coupled IR Banks-Zaks fixed point~\cite{Banks:1981nn}, insufficient to generate large
anomalous dimensions. On the other hand, the lower edge of the IR conformal window, below which the
theory condenses, is expected to lie at $N_F\approx 10$ (using the Schwinger-Dyson equation~\cite{Sannino:2009za}) or
$N_F\approx 4$ (using the Pica-Sannino beta function~\cite{Pica:2010mt}). Thus, it is convenient to have
as few $\tbf{F}$-hyperfermions as possible at low energy. We will therefore assume that $L_u^p,L_d^p,U_d^p,D_u^p$, $(p=1,2)$,
acquire a large mass via a hierarchy $\lambda_{\Theta L/R}^{11,1} \sim \lambda_{\Theta L/R}^{22,1} \sim \mathcal{O} (1)\ll \lambda_{\Theta L/R}^{33,1}$.
In this scenario, the EWSB is thus due to the hyperfermions associated to the third generation fermions.
Below the scale $\Lambda_F$, therefore, the theory contains $N_F = 12$.
At this stage, only a non-perturbative lattice calculation can prove if this theory allows for large enough anomalous dimensions for the PC operators~\cite{Lee:2018ztv}.

Furthermore, the mass generation for the 1$^{\rm st}$ and 2$^{\rm nd}$ families entails the breaking of some accidental global symmetries present in our model: a global $SU(2)_{Lp}$ rotation of each $\Omega^p$ multiplet individually, and a $\mathbb{Z}_{2p}$ parity under which $\Omega^p$ is odd.
Without explicit breaking via family-changing Yukawas, the only possibility would be breaking them spontaneously, leading to violation of the 't Hooft decoupling condition. It is, therefore, compulsory to introduce the $\Theta^2$ field to enable the light families to feel the EWSB.
Via a mixing of the scalar components in $\Theta^2$ and $\Psi^2$, a flavor structure proportional to $\lambda_{\Theta L/R}^{p3,2} \lambda_\Psi^2$ will be generated. Together with the gauge mediation, this will give mass to only two generations: masses for a third one can be generated by the $\Delta$--mediators, proportional to $\lambda_{\Delta L/R}^p \lambda_{\Delta L/R}^3$. Another possibility is that box loops containing the scalars in $\Theta^\alpha$--$\Psi^\alpha$ and the heavy 1$^{\rm st}$ and 2$^{\rm nd}$ generation hyperfermions could generate sizeable PC couplings proportional to a linear combination of $\lambda_{\Theta L/R}^{p2,2}$ and  $\lambda_{\Theta L/R}^{p1,2}$. In the loop-generated case, there is no need to introduce the $\Delta$ scalars. More details will be presented in ref.~\cite{Giacomo:2019ehd}.

In this model, baryon ($B$), lepton ($L$) and hyperbaryon ($H_B$) numbers can be defined in terms of the gauged $U(1)$'s, namely $U(1)_E$ and $U(1)_X$ from the PU breaking, and of the global symmetry $U(1)_G$. With standard assignment for the first two, and $H_B (L^p_{u/d}) = 1/2$, we find $Q_G = 3B + L + 2 H_B$. These symmetries are preserved by the Yukawa couplings in eq.~\eqref{eq:Yuk} and by gauge interaction. However, $L$ is broken by $v_\Phi$, like in the standard Pati-Salam, while $H_B$ is broken by $v_\Theta$, once the complex hypercolor group is broken. In the scenario we presented in this letter, therefore, baryon number $B$ is preserved, avoiding the strong bounds from proton and neutron decays. As a consequence, the components of $\Xi$ have exotic $B$ assignments: for instance, the singlet $\rho$ has $B=1/2$, thus forbidding any decay into SM states. This state could therefore be an accidental Dark Matter candidate in the model, with a mass around the condensation scale as it is linked to the masses of $\chi$ and $\eta$ via $\mu_\Xi$.
Small $B$-violation effects could be present in the scalar potential $V(\Phi,\Theta,\Delta,\Psi)$ via $U(1)_G$ breaking couplings like $\Theta^\alpha \Delta_{L/R} \Delta_{L/R}$ trilinears, and their effect can be controlled via a small coupling constant. An interesting variation of the model could arise from a vacuum expectation value acquired by components of $\Delta_R$, which could be responsible for the breaking steps {\it b} and {\it c}. Via $\lambda_{\Delta R}^3$, a mixing between $b_R^c$--$\tilde{\omega}$ and between the hyperfermions $D^3_u$--$U_u$ and $U^3_d$--$D_d$ is generated: either one affects the couplings of $b_R^c$ to the PC operators, thus potentially explaining the $t$--$b$ mass splitting. The viability of this scenario, however, can only be tested once the conformal dynamics is known, as $B$-violating effects could be enhanced by large anomalous dimensions.

To conclude, we consider \emph{partial unification} as an attractive approach to construct UV complete composite Higgs models with fermion partial compositeness.
We presented a first attempt to build a fully realistic model based on a Pati-Salam inspired unification and renormalizable
 up to the Planck scale. At low energy, we predict a composite dynamics based on $Sp(4)_{\rm HC}$ hypercolor
 with $6\times\textbf{A}_2$ and $12\times\textbf{F}$ Weyl
flavors. Lattice validation is needed to check if such model does approach a strongly-coupled IR Banks-Zaks fixed point,
necessary to generate large anomalous dimensions. The theory then exits the walking regime at a lower scale, $\Lambda_{\rm HC} \approx 10$~TeV,
where the Higgs coset is $SU(4)/Sp(4)$ or $SU(6)/Sp(6)$ depending on the hyperfermion masses.
In this letter we have demonstrated that fermion masses and flavor mixing can be generated, while
more detailed aspects will be presented in a forthcoming publication~\cite{Giacomo:2019ehd}.
We focus on a scenario where baryon number is preserved, thus avoiding proton/neutron decays, offering as
a consequence an accidental fermionic Dark Matter candidate around the TeV scale.
The results presented in this letter offer a novel perspective on composite dynamics for the Higgs, and open up
a new direction in studying UV completions valid up to the Planck scale. The most interesting prediction is the
dynamics at low energy, which can be studied on the lattice. In fact, lattice results in the conformal window are
needed in order to further study the flavor structure at low energy and test if flavor bounds are avoided.


\subsection*{Acknowledgements}

CZ would like to thank Kingman Cheung and Jean-Pierre Derendinger  for helpful discussions.
GC and SV received partial support from the Labex-LIO (Lyon Institute of Origins) under grant ANR-10-LABX-66 (Agence Nationale pour la Recherche), and FRAMA (FR3127, F\'ed\'eration de Recherche ``Andr\'e Marie Amp\`ere'').
We also thank the Sun-Yat Sen University in Guangzhou for hospitality, where this project was initiated.


\bibliography{tps}
\bibliographystyle{h-physrev}

\end{document}